%% file: lse_precoding_general_replica.tex
\documentclass[conference]{IEEEtran}
\ifCLASSINFOpdf
\else
\fi
\usepackage{amsmath,dsfont,bbm,epsfig,amssymb,amsfonts,amstext,verbatim,amsopn,cite,subfigure,multirow,multicol,lipsum,xfrac}
\usepackage{amsthm}
\usepackage{mathtools,amsthm}
\usepackage{perpage}
\usepackage{balance}
\usepackage{url}
\usepackage{amsfonts}
\usepackage{epsfig}
\usepackage[font={small}]{caption}
\usepackage{psfrag}	
\usepackage{etoolbox}
\usepackage{algorithmicx}
\usepackage[Algorithm,ruled]{algorithm}
\usepackage{algpseudocode}
\usepackage{pifont}
\usepackage[utf8]{inputenc}
\usepackage[T1]{fontenc}  
\usepackage[nolist]{acronym}
\MakePerPage{footnote}
\usepackage{paralist}
\usepackage{enumitem}
\usepackage{bbm}
\usepackage[process=auto]{pstool}
\usepackage{tikz}
\usetikzlibrary{shapes,arrows}
\include{commands}

\begin{document}
%
\title{Asymptotics of Nonlinear LSE Precoders with Applications to Transmit Antenna Selection\vspace{-2.5mm}}

\author{
\IEEEauthorblockN{
Ali Bereyhi, 
Mohammad Ali Sedaghat, 
Ralf R. M\"uller
}
\IEEEauthorblockA{
Institute for Digital Communications (IDC), Friedrich-Alexander Universit\"at Erlangen-N\"urnberg (FAU)\\
ali.bereyhi@fau.de, mohammad.sedaghat@fau.de, ralf.r.mueller@fau.de \vspace*{-6mm}
\thanks{This work was supported by the German Research Foundation, Deutsche Forschungsgemeinschaft (DFG), under Grant No. MU 3735/2-1.}
}
}


%


\IEEEoverridecommandlockouts

\maketitle

\begin{abstract}
This paper studies the large-system performance of Least Square Error (LSE) precoders which~minimize~the~input-output distortion over an arbitrary support subject to a general penalty function. The asymptotics are determined via the replica method in a general form which encloses the Replica Symmetric (RS) and Replica Symmetry Breaking (RSB) ans\"atze. As a result, the ``marginal decoupling property'' of LSE precoders for $b$-steps of RSB is derived. The generality of the studied setup enables us to address special cases in which the number of active transmit antennas are constrained. Our numerical investigations depict that the computationally efficient forms of LSE precoders based on ``$\ell_1$-norm'' minimization perform close to the cases with ``zero-norm'' penalty function which have a considerable improvements compared to the random antenna selection. For the case with BPSK signals and restricted number of active antennas, the results show that RS fails to predict the performance while the RSB ansatz is consistent with theoretical bounds.
\end{abstract}

\IEEEpeerreviewmaketitle

\section{Introduction}
For the \ac{mimo} channel
\begin{align}
\by=\mH \hspace*{.5mm} \bx + \bz \label{eq:ch}
\end{align}
with $\mH \in \setC^{k\times n}$, $\bx \in \setX^n$ and $\bz\sim \mathcal{CN}(\boldsymbol{0},\lambda_z\mI_k)$, the nonlinear \ac{lse} precoder with the general penalty function $u(\cdot)$ is given by
\begin{align}
\bx=\arg \min_{\bv \in {\setX^n}} \norm{\mH \bv-\sqrt{\rho} \hspace*{.5mm}\bs}^2 +u(\bv). \label{eq:1}
\end{align}
The precoder maps the $k$-dimensional source vector $\bs$, scaled with the power control factor $\rho$, to the $n$-dimensional input vector $\bx$ whose entries are taken from the given support $\setX$. The mapping is such that the distortion caused by the channel impact, i.e., $\norm{\mH \hspace*{.5mm}\bx-\sqrt{\rho} \hspace*{.5mm}\bs}^2$, is minimized over the given input support $\setX^n$ subject to some constraints imposed by $u(\cdot)$. The conventional precoding schemes such as \ac{rzf}, Tomlinson-Harashima or vector precoding, mostly consider the average transmit power constraint and assume the set of possible input constellation points to be the complex plane, i.e., $\setX=\setC$. The latter consideration was partially relaxed in \cite{mohammed2013per} where authors studied the ``per-antenna constant envelope precoding''. The set of possible constellation points was later generalized to an arbitrary set by introducing a class of power-limited nonlinear precoders \cite{sedaghat2017lse}. The precoder in \eqref{eq:1} generalizes the earlier schemes by letting different types of constraints be imposed on the precoded vector. In fact, due to the generality of the penalty function the scope of restrictions on $\bx$ is broaden.
Consequently, several precoding schemes are considered as special cases of \eqref{eq:1}. To name some examples, let $u(\bv)=\lambda \norm{\bv}^2$; then, for $\setX=\setC$, the precoder reduces to the~\ac{rzf}~precoder introduced in \cite{peel2005vector}, and by considering $\setX=\set{v\in\setC: \abs{v}=\sfK}$ for some constant $\sfK$, the precoder reduces to a constant envelope precoder \cite{mohammed2013per}.

This paper investigates the asymptotic performance of the precoder. Our motivation comes from recent promising results reported for massive \ac{mimo} systems  \cite{hoydis2013massive}. For some choices of $\setX$ and $u(\cdot)$, the system can be asymptotically analyzed via tools from random matrix theory \cite{schmidt2008minimum}. The tools, however, fail to study the large-system performance of the precoder for many other choices. Therefore, we invoke the ``replica method'' developed in statistical mechanics. In the context of multiuser systems, the replica method was initially utilized by Tanaka in \cite{tanaka2002statistical} to study the asymptotic performance of randomly spread CDMA detectors. The method was later  widely employed for large-system analysis in communications and information theory; see for example \cite{zaidel2012vector} and the references therein.
\subsection*{Contributions}
For nonlinear \ac{lse} precoders, we determine the input-output distortion, as well as the marginal distribution of output~entries, in the large-system limit via the replica method. We deviate from our earlier replica symmetric study in \cite{bereyhi2017nonlinear}, by determining the general replica ansatz which includes both the replica symmetry and symmetry breaking ans\"atze. Our general result furthermore depicts that under any assumed replicas' structure, the output symbols of the precoder marginally decouple~in~the asymptotic regime. A brief introduction to the replica method is given in the appendix through the large-system analysis. As an application, we study special cases of the precoder with~co- nstraints on the number of active antennas. Our numerical~inv- estigations show that computationally efficient \ac{lse} precoders based on $\ell_1$-norm minimization perform significantly close to \ac{lse} precoders with zero-norm penalty. Moreover, the problem of BPSK transmission with constraint on the number of active antennas is shown to exhibit replica symmetry~breaking.
\subsection*{Notation}
We represent scalars, vectors and matrices with non-bold, bold lower case and bold upper case letters, respectively. A $k \times k$ identity matrix is shown by $\mI_k$, and the $k \times k$ matrix with all entries equal to one is denoted by $\mone_k$. $\mH^{\her}$ indicates the Hermitian of the matrix $\mH$. The set of real and integer numbers are denoted by $\setR$ and $\setZ$, and their corresponding non-negative subsets by superscript $+$; moreover, $\setC$ represents the complex plane. For $s\in\setC$, $\re{s}$  and $\sphericalangle s$ identify the real part and argument, respectively. $\norm{\cdot}$ and $\norm{\cdot}_1$ denote the Euclidean and $\ell_1$-norm, respectively, and $\norm{\bx}_0$ represents the zero-norm defined as the number of nonzero entries. For a random variable $x$, $\mathrm{p}_x$ represents either the probability mass or density function. Moreover, $\E$ identifies the expectation operator. For sake of compactness, the set of integers $\set{1, \ldots , n}$ is abbreviated as $[1:n]$ and a zero-mean complex Gaussian distribution with variance $\rho$ is represented by $\phi(\cdot;\rho)$. Whenever needed, we assume the support $\setX$ to be discrete. The results, however, are in full generality and hold also for continuous distributions.

\section{Problem Formulation}
\label{sec:sys}
Consider the precoding scheme illustrated in \eqref{eq:1} in which 
\begin{enumerate}[label=(\alph*)]
\item $\mH_{k\times n}$ is a random matrix whose eigendecomposition is $\mH^{\her} \mH = \mU \mD \mU^{\her}$ with $\mU_{n \times n}$ being a Haar distributed unitary matrix, and $\mD_{n \times n}$ being a diagonal matrix with asymptotic eigenvalue distribution $\rmp_\mD$.
\item $\bs_{k \times 1}$ has \ac{iid} zero-mean and unit-variance complex Gaussian entries, i.e., $\bs\sim\mathcal{CN}(\boldsymbol{0}, \mI_{k})$ and is independent of $\mH$.
\item $\rho$ is a non-negative real power control factor.
\item $u(\cdot)$ is a general penalty function with decoupling property, i.e., $u(\bv)=\sum_{j=1}^n u(v_j)$.
\item The dimensions of $\mH$ grow large, such that the~load~factor, defined as $\alpha\coloneqq k/n$, is kept fixed in both $k$~and~$n$.
\end{enumerate}
%
%
%
%
%
%
%
For this setup, we define the asymptotic marginal as follows.
\begin{definition}[\bfseries Asymptotic Marginal]
\label{def:margin}
Consider the function $f(\cdot): \setX \mapsto \setR$. The marginal of $f(\bx)$ over $\setW(n) \subseteq [1:n]$ is
\begin{align}
\sfM_{f}^{\setW}(\bx;n)\coloneqq \frac{1}{\abs{\setW(n)}}\hspace*{1mm} \E \hspace*{-2.1mm}\sum_{w\in\setW(n)}  f(x_w)
\end{align}
The asymptotic marginal of $f(\bx)$ is then defined to be the limit of $\sfM^{\setW}_f(\bv;n)$ as $n\uparrow\infty$, i.e., $\sfM^{\setW}_f(\bx) \coloneqq \lim\limits_{n\uparrow\infty}\sfM^{\setW}_f(\bx;n)$.
\end{definition}
The asymptotic marginal of $f(\bx)$ determines large-system characteristics of $\bx$ including the marginal distribution of its entries. In order to quantify the large-system performance, we further define the asymptotic distortion as a measure.
\begin{definition}[\bfseries Asymptotic Distortion]
\label{def:asy_dist}
For the precoder given in \eqref{eq:1}, the asymptotic input-output distortion is defined~as
\begin{align}
\sfD (\rho) \coloneqq \lim_{k\uparrow\infty} \frac{1}{k} \E \norm{\mH \bx-\sqrt{\rho}\hspace*{.5mm} \bs}^2.
\end{align}
\end{definition}
\section{Main Results}
\label{sec:result}
We start by defining the $\rmR$-transform of a distribution.
\begin{definition}[\bfseries $\rmR$-transform]
\label{def:r-trans}
For $t$ with distribution $\rmp_t$, the Stieltjes transform over the upper complex half plane is given by $\rmG_t(s)\hspace*{-.7mm}= \hspace*{-.7mm}\E (t-s)^{-1}$. Denoting the inverse \ac{wrt} composition by $\rmG_t^{-1} (\cdot)$, the {$\rmR$-transform} of $\rmp_t$ is defined as $\rmR_t (\omega) \hspace*{-.7mm}=\hspace*{-.7mm} \rmG_t^{-1} (-\omega) - \omega^{-1}$ such that $\lim\limits_{\omega\downarrow 0} \rmR_t (\omega)\hspace*{-.7mm} =\hspace*{-.7mm} \E t$.\\ Moreover, let $\mM_{n \times n}$ be decomposed as $\mM=\mU \mathbf{\Lambda} \mU^{-1}$ where $\mathbf{\Lambda}_{n \times n}$ is the diagonal matrix of eigenvalues, and $\mU_{n \times n}$ is the matrix of eigenvectors. Then $\rmR_t(\mM)$ is an $n \times n$ matrix defined as $\mathrm{R}_t(\mM)=\mU \ \mathrm{diag}[\mathrm{R}_t(\lambda_1), \ldots, \mathrm{R}_t(\lambda_n)] \ \mU^{-1}$.
\end{definition}
Proposition \ref{thm:1} expresses $\sfM^{\setW}_f(\bx)$ and $\sfD(\rho)$ in terms of the $\rmR$-transform of $\rmp_\mD$. The result is determined for a general structure of replicas, and only relies on the replica continuity assumption which is briefly explained in the appendix.
\begin{proposition}[\bfseries General Replica Ansatz]
\label{thm:1}
Consider the nonlinear \ac{lse} precoder in Section \ref{sec:sys}, and define $\bvv_{m\times1}$ to be a random vector over $\setX^m$ with the~distribution~$\rmp_{\bvv}^\beta(\bvv;\mQ)$ 
\begin{align}
\rmp_{\bvv}^\beta(\bvv;\mQ)&=\dfrac{e^{-\beta\left[\bvv^\her \mT \rmR_{\mD} (-\beta \mT \mQ) \bvv+ u(\bvv)\right] }}{\sum_{\bvv} e^{-\beta\left[\bvv^\her \mT \rmR_{\mD} (-\beta \mT \mQ) \bvv+ u(\bvv)\right] }}. 
\end{align}
for some $m\times m$ matrix $\mQ$ with real entries, non-negative real scalar $\beta$, and $\mT \coloneqq \mI_m- \dfrac{\beta\rho}{1+m\beta\rho} \mone_m$. 
%
Let $\mQ^\star$ satisfy 
\begin{align}
\mQ^\star&=\sum_{\bvv} \rmp_{\bvv}^\beta(\bvv;\mQ^\star) \bvv \bvv^\her. \label{eq:33}
\end{align}
Then, under the replica continuity assumption, the asymptotic marginal of $f(\bx)$ is given by
\begin{align}
\sfM^{\setW}_f(\bx)=\lim_{\beta\uparrow\infty} \lim_{m\downarrow0} \sum_{\bvv} \rmp_{\bvv}^\beta(\bvv;\mQ) \sfM^{\setT}_f(\bvv;m), \label{eq:35}
\end{align}
and $\sfD (\rho) = \rho + {\alpha}^{-1} \lim\limits_{\beta\uparrow\infty}  \mad^\sfR(\beta)$ where $\mad^\sfR(\cdot)$ is defined~as
\begin{align}
\mad^\sfR(\beta) \coloneqq  &\frac{\partial}{\partial \beta} \left[ \lim_{m\downarrow 0} \frac{1}{m} \tr{\int_0^\beta \mT \mQ^\star \rmR_{\mD}(-\omega \mT\mQ^\star) \dif\omega} \right] \nonumber\\
&-\beta \lim_{m\downarrow 0} \frac{1}{m}  \tr{\mT \rmR_\mD(-\beta\mT\mQ^\star) \frac{\partial\mQ^\star}{\partial \beta}} .
\end{align}
\end{proposition}
\begin{prf}
The proof is briefly addressed in the appendix.~The~details, however, are omitted due to lack of space and will~be~fo- rthcoming in the extended version of the paper.
\end{prf}
To determine $\sfM^{\setW}_f(\bx)$ and $\sfD (\rho)$ in Proposition \ref{thm:1}, one needs to determine the fixed-point $\mQ^\star$ through \eqref{eq:33}, and then, find the function at the \ac{rhs} of \eqref{eq:35} and $\mad^\sfR(\beta)$ in an analytic form. Finding the solution of \eqref{eq:33}, however, is notoriously difficult and possibly some of the solutions are not of use. The trivial approach is to restrict the search to a set of parameterized matrices. The most primary set is given by \ac{rs}. The \ac{rs} solution, however, may result in an invalid prediction of the performance. A more general structure is given by imposing the \ac{rsb} structure which we address in the sequel. 
\subsection{General Marginal Decoupling Property}
Proposition \ref{thm:1} enables us to investigate a more general form of the ``asymptotic marginal decoupling property'' introduced in \cite{bereyhi2017nonlinear}. The property indicates that in the large-system limit, the marginal distribution of all output entries are identical and expressed as the output distribution of an equivalent single-user system. In fact, it can be considered as~a~dual~version~of the decoupling property investigated in the literature for different classes of nonlinear estimators, e.g. \cite{guo2005randomly, rangan2012asymptotic, bereyhi2016rsb}. As the analysis in \cite{bereyhi2017nonlinear} was under the \ac{rs} assumption, the result was limited to the cases in which \ac{rs} assumption gives a valid prediction. The generality of Proposition \ref{thm:1}, however, enables us to investigate this property of the precoder for any structure of replicas. To illustrate the property, consider the following definition.
\begin{definition}
Denote the marginal distribution of the $j$th entry of $\bx_{n\times1}$, i.e., $x_j$ for some $j\in [1:n]$, by $\rmp^{j(n)}_x$ where the superscript $n$ indicates the dependency on the length of $\bx$. Then, the asymptotic marginal distribution $\rmp^{j}_x$ is defined to be the limit of $\rmp^{j(n)}_x$ as $n\uparrow\infty$, i.e., $\rmp^{j}_x(t)\coloneqq\lim\limits_{n\uparrow\infty} \rmp^{j(n)}_x(t)$.
\end{definition}
\begin{decoupling}
\label{thm:2}
Consider the nonlinear \ac{lse} precoder with the constraints given in Section~\ref{sec:sys}. Then, under the replica continuity assumption, the asymptotic marginal distribution $\rmp^{j}_x$ converges to a deterministic distribution which is constant in $j$ for any $j\in [1:n]$ regardless of the structure imposed on $\mQ^\star$.
\end{decoupling}
\subsection{\ac{rsb} Ans\"atze}
Parisi proposed the method of \ac{rsb} to construct a set of parameterized matrices which recursively extends to larger classes. The method starts from the \ac{rs} structure for $\mQ^\star$, and then recursively constructs new structures. After $b$ steps of recursion, $\mQ^\star$ becomes of the form
\begin{align}
\mQ^\star= \frac{\chi}{\beta} \mI_m + \sum_{\kappa=1}^b \sfc_\kappa\hspace*{.5mm} \mI_{\frac{m\beta}{\mu_\kappa}} \otimes \mone_{\frac{\mu_\kappa}{\beta}} +  \sfp \mone_m, \label{eq:rsb}
\end{align}
for some non-negative real scalars $\chi$, $\beta$ and $\sfp$, and sequences $\set{\sfc_\kappa}$ and $\set{\mu_\kappa}$. The structure in \eqref{eq:rsb} reduces to \ac{rs} by setting $\set{\sfc_\kappa}\equiv 0$. By substituting \eqref{eq:rsb} in Proposition~\ref{thm:1}, the $b$-steps \ac{rsb} ansatz is determined. For cases that the \ac{rs} ansatz gives the exact solution, the coefficients $\set{\sfc_\kappa}$ at the saddle points are equal to zero. However, in cases that \ac{rs} fails, the sequence $\set{\sfc_\kappa}$ has non-zero entries. The investigations in \cite{sedaghat2017lse} show that the \ac{rs} ansatz clearly fails giving a valid prediction of the performance in some cases. Therefore, the \ac{rsb} ans\"atze are required to be considered further. For sake of compactness, we state the one-step \ac{rsb} ansatz, i.e., $b=1$, in this paper. The result, however, is extended to an arbitrary number of breaking steps by taking the approach in Appendix D of \cite{bereyhi2016statistical}. 


%
%
%
\begin{corollary}[\bfseries One-step \ac{rsb} Ansatz]
\label{thm:2}
Let the assumptions in Proposition \ref{thm:1} hold, and consider $\mQ^\star$ to be of the form \eqref{eq:rsb} with $b=1$. For given $\chi$, $\sfp$, $\mu$ and $\sfc$, define $\rho^\rs$ and $\rho^\rsb_1$ as
\begin{subequations}
\begin{align}
\rho^\rs&=\xi^{2}\frac{\partial}{\partial \tilde{\chi}} \left[ ( \rho \tilde{\chi}- \sfp ) \rmR_\mD(-\tilde{\chi}) \right]\\
\rho^\rsb_1&=\xi^{2} \mu^{-1} \left[ \rmR_\mD(-\chi)-\rmR_\mD(-\tilde{\chi}) \right] 
\end{align}
\end{subequations}
where $\tilde{\chi}\coloneqq\chi+\mu\sfc$ and $\xi\coloneqq[\rmR_\mD(-\chi)]^{-1}$. Let $\xx$ be
\begin{align}
\xx=\arg \min_v \abs{v-  s^\rs - s^\rsb_1}^2+ \xi\hspace*{.5mm} u(v). \label{eq:single}
\end{align}
where $s^\rs\hspace*{-.7mm}\sim\hspace*{-.7mm}\phi(\cdot;\rho^\rs)$, and $s^\rsb_1$ is obtained by passing $s^\rs$ through 
\begin{align}
\hspace*{-1mm}\rmp_1^\rsb(u|t) \hspace*{-.7mm}=\hspace*{-.7mm}\frac{e^{-\tfrac{\mu}{\xi} \left[ \abs{\xx-  u - t}^2-\abs{ u + t}^2 \right] -\mu u(\xx)} \phi(u;\rho_1^\rsb)}{\int_{\setC} e^{-\tfrac{\mu}{\xi} \left[ \abs{\xx-  w - t}^2-\abs{ w + t}^2 \right] -\mu u(\xx)} \phi(w;\rho_1^\rsb) \dif w} \label{final_prob}
\end{align}
%
Then, $\sfM^{\setW}_f(\bx)= \E f(\xx)$, and the asymptotic distortion reads
\begin{align}
\hspace*{-1.6mm}\sfD(\rho)\hspace*{-.7mm}=\hspace*{-.7mm}\rho \hspace*{-.7mm} +\hspace*{-.7mm} \alpha^{-1} \left\lbrace \frac{\partial}{\partial \tilde{\chi}}\hspace*{-.7mm} \left[ ( \sfp\hspace*{-.7mm} -\hspace*{-.7mm}\rho \tilde{\chi} ) \tilde{\chi} \rmR_\mD(-\tilde{\chi}) \right]  \hspace*{-.7mm}+\hspace*{-.7mm} \frac{\xi\sfp \hspace*{-.7mm} - \hspace*{-.7mm} \tilde{\chi}\rho_1^\rsb}{\xi^2} \right\rbrace.\label{final_dist}
\end{align}
In \eqref{final_prob} and \eqref{final_dist}, ${\chi}$, $\sfc$ and $\sfp$ are determined via the equations
\begin{subequations}
\begin{align}
\sfc+\sfp&=\E \abs{\xx}^2\\
\sfp+\tilde{\chi} &= \frac{\xi}{\rho_1^\rsb} \hspace*{.5mm} \E \re{\xx^* s_1^\rsb}\\
\tilde{\chi} &= \frac{\xi}{\rho^\rs} \hspace*{.5mm} \E \re{\xx^* s^\rs}.
\end{align}
\end{subequations}
and $\mu$ satisfies the following fixed-point equation
\begin{align}
\hspace*{-1mm}\frac{\mu^2 \sfp}{\xi^2} \rho_1^\rsb + \frac{\mu \sfc}{\xi} +\mai \hspace*{-.7mm}=\hspace*{-.7mm} \mathrm{I} \left( s_1^\rsb;s^\rs \right) + \mathrm{D}_{\mathsf{KL}} ( \rmp_{s_1^\rsb} \Vert \phi(\cdot;\rho_1^\rsb) ) \label{eq:fix3}
\end{align}
where $\rmp_{s_1^\rsb}(u)= \int \rmp_1^\rsb (u|t) \phi(t;\rho^\rs) \dif t$, $\mathrm{D}_{\mathsf{KL}} ( \cdot \Vert \cdot )$ denotes the Kullback–Leibler divergence, and $\mai\coloneqq-\int_{\chi}^{\tilde{\chi}} \rmR_{\mD}(-\omega) \dif \omega$.
\end{corollary}
\begin{remark}
The ansatz in Corollary \ref{thm:2} reduces to \ac{rs} \cite{bereyhi2017nonlinear}, by enforcing the fixed-point solution to have $\sfc=0$. The \ac{rs}~ansatz, however, is not necessarily valid. The valid solution here is chosen such that the corresponding free energy is minimized. 
\end{remark}
\begin{decouplersb}
Considering the one-step \ac{rsb} ansatz, the asymptotic marginal distributions of the precoded symbols are described by $\xx$; more precisely, for any $j\in [1:n]$ we have $\rmp^j_x\equiv\rmp_\xx$. The distribution can be described by an equivalent single-user system which we refer to as the ``decoupled precoder'', and is defined as
\begin{align}
\xx^\dec (s^\dec)=\arg \min_v \abs{v-  s^\dec}^2+ \xi \hspace*{.5mm} u(v). \label{eq:single_final}
\end{align}
The one-step \ac{rsb} decoupled precoder is similar to \ac{rs};~however, the ``decoupled input'' $s^\dec$, which in \ac{rs} is $s^\rs$, is replaced by $s^\rs + s^\rsb_1$. Taking the same approach as in \cite{bereyhi2016statistical}, it is shown that under $b$-steps of \ac{rsb}, the decoupled precoder has a same form, and $s^\dec=s^\rs + \sum_{\kappa=1}^b s^\rsb_\kappa$. In this case, $s^\rsb_\kappa$ is obtained from $s^\rs$ and $\set{s^\rsb_\varsigma}_{\varsigma=\kappa+1}^b$ through $\rmp^\rsb_\kappa (u_\kappa|u_{\kappa+1}, \ldots,u_b , t)$. 
\end{decouplersb}
\section{Applications to Transmit Antenna Selection}
\label{sec:app}
As we discussed, considering a general penalty function lets us investigate several transmit constraints. Restrictions on the number of active antennas is a constraint which arises in \ac{mimo} systems with \ac{tas} \cite{molisch2005capacity}. The goal in these systems is to minimize the number of \ac{rf} chains which significantly reduces the overall \ac{rf}-cost. The fundamental limits as well as efficient selection algorithms, however, have not been yet precisely addressed in the literature. In this section, we investigate the asymptotics of some special cases of the \ac{lse} precoder which imply \ac{tas}.


%
\subsection{\ac{tas} by Zero-Norm Minimization}
\label{sec:ex1}
The \ac{lse} precoder with $u(\bv)=\lambda \norm{\bv}^2 + \lambda_0 \norm{\bv}_0$ imposes constraints on the average transmit power and number of active antennas. For $\setX=\setC$, the decoupled precoder reads
\begin{align}
\xx^\dec(s^\dec)=
\begin{cases}
    \dfrac{s^\dec}{1+\xi\lambda} \qquad &\abs{s^\dec}\geq \tau_0 \\
    0             &\abs{s^\dec} < \tau_0 \label{eq:sing0}
\end{cases}
\end{align}
for $\tau_0\coloneqq\sqrt{\xi\lambda_0 (1+\xi\lambda)}$. Here, the decoupled precoder is a hard thresholding operator. As $\lambda_0\downarrow 0$, $\tau_0$ tends to zero as well. For the case with limited peak power where for some $\rmP\in\setR^+$
\begin{align}
\setX = \set{ r e^{\mathrm{j} \theta }: \hspace*{2mm} 0 \leq \theta \leq 2\pi \ \wedge \ 0 \leq r \leq \sqrt{\rmP}},
\end{align}
the decoupled precoder is given by
\begin{align}
\xx^\dec(s^\dec)=
\begin{cases}
  \dfrac{ s^\dec}{\abs{s^\dec}} \sqrt{\rmP} \qquad  & \hat{\tau}_0 \leq \abs{s^\dec} \\
    0             & \tilde{\tau}_0 \leq \abs{s^\dec} < \hat{\tau}_0 \\
    \dfrac{s^\dec}{1+\xi\lambda} &\tau_0 \leq \abs{s^\dec}\leq \tilde{\tau}_0 \\
    0             &\hspace*{.8mm}0\hspace*{.3mm}\leq \abs{s^\dec} < \tau_0 \label{eq:single1}
\end{cases}
\end{align}
where $\tilde{\tau}_0=(1+\xi\lambda)\sqrt{\rmP}$ and $\hat{\tau}_0=\max\set{\tilde{\tau}_0, \tilde{\tau}_0/2+\tau_0^2/2\tilde{\tau}_0}$. The decoupled precoder in \eqref{eq:single1} is a two-steps hard thresholding operator which in the first step constrains the transmit peak power, and in the second step, implies the \ac{tas} constraint. By setting $\lambda_0=0$, $\tau_0$ becomes zero and $\hat{\tau}_0=\tilde{\tau}_0$.

The \ac{lse} precoders with zero-norm penalty function need~to minimize a non-convex function which has a high computational complexity. We therefore propose~an alternative form of the precoder based on the $\ell_1$-norm minimization.
\subsection{\ac{tas} by $\ell_1$-Norm Minimization}
\label{sec:ex2}
To reduce the complexity of the precoding~schemes~in~Section~\ref{sec:ex1}, we modify $u(\cdot)$ as $u(\bv)=\lambda \norm{\bv}^2 + \lambda_1 \norm{\bv}_1$. The objective function in this case is convex, and therefore, for convex choices of $\setX$, the resulting form of the \ac{lse} precoder is effectively implemented by employing computationally feasible algorithms. We start by considering $\setX=\setC$ in which
\begin{align}
\xx^\dec(s^\dec)=
\begin{cases}
   \dfrac{s^\dec}{1+\xi\lambda} \dfrac{\abs{s^\dec}-\tau_1}{\abs{s^\dec}} &\abs{s^\dec}\geq \tau_1 \\
    0             &\abs{s^\dec} < \tau_1 \label{eq:single_norm1}
\end{cases}
\end{align}
with $\tau_1\coloneqq {\xi \lambda_1}/{2}$. The decoupled precoder in this case is a soft thresholding operator. In fact, \eqref{eq:single_norm1} is obtained from \eqref{eq:sing0} by multiplying the factor $1-\tau_1/{\abs{s^\dec}}$. Similar to \eqref{eq:sing0}, the threshold in \eqref{eq:single_norm1} tends to zero as $\lambda_1\downarrow 0$. For the case with limited peak transmit power, the decoupled precoder reads
\begin{align}
\hspace*{-1mm}\xx^\dec(s^\dec)=
\begin{cases}
 \dfrac{s^\dec}{\abs{s^\dec}} \sqrt{\rmP} &\tilde{\tau}_1\leq \abs{s^\dec}\\
    \dfrac{s^\dec}{1+\xi\lambda} \dfrac{\abs{s^\dec}-\tau_1}{\abs{s^\dec}}   &\tau_1 \leq \abs{s^\dec} < \tilde{\tau}_1 \\
    0             &\hspace*{.8mm}0\hspace*{.3mm}\leq \abs{s^\dec} < \tau_1 \label{eq:single_norm11}
\end{cases}
\end{align}
for $\tau_1\coloneqq {\xi \lambda_1}/{2}$ and $\tilde{\tau}_1 \coloneqq \sqrt{\rmP}(1+\xi\lambda)+{\xi \lambda_1}/{2}$. As in \eqref{eq:single1}, the decoupled precoder in \eqref{eq:single_norm11} is a two-steps thresholding. In the first step, $s^\dec$ is constrained \ac{wrt} the peak power $\rmP$ via a hard thresholding operator with level $\tilde{\tau}_1$, and then at the second step, the \ac{tas} constraint is imposed on the decoupled input by a soft thresholding operator as in \eqref{eq:single_norm1}. By setting $\lambda_1=0$, the threshold $\tau_1$ reads $\tau_1=0$ and $\tilde{\tau}_1=\sqrt{\rmP}(1+\xi\lambda)$.
\subsection{\ac{tas} with $\sfM$-PSK Signals on Antennas}
\label{sec:ex3}
Considering the precoding support as $\setX = \{ 0, \sqrt{\rmP} \hspace{.2mm} e^{\rmj\tfrac{2k \pi}{\sfM}}\}$, for $k\in [1:\sfM]$, the precoder is constrained to map the source to a vector of $\sfM$-PSK symbols over a subset of antennas while keeping the others silent. In this case, the transmit power on each active antenna is $\rmP$, and therefore, $\norm{\bx}^2=\rmP \norm{\bx}_0$ which indicates that any restriction on the average transmit power imposes a proportional constraint on the number of active antennas. Consequently, \ac{tas} is applied via the \ac{lse}~precoder by setting the penalty function as $u(\bv)=\lambda\norm{\bv}^2$. By defining the function $\psi(\cdot)$ as $\psi(k)\coloneqq \cos\left( \frac{2k\pi}{\sfM}-\sphericalangle s^\dec \right)$, the decoupled precoder in this case is derived as
\begin{align}
\xx^\dec(s^\dec)=
\begin{cases}
     \sqrt{\rmP} \hspace{.2mm} e^{\rmj\tfrac{2k^\star \pi}{\sfM}} &\abs{s^\dec}\geq \tau_\sfd \\
    0             &\abs{s^\dec} < \tau_\sfd \label{eq:single_psk}
\end{cases}
\end{align}
where $\tau_\sfd\coloneqq\sqrt{\rmP} (1+\xi\lambda)\psi(k^\star)^{-1}/2$ for $k^\star \coloneqq \arg\max_{k} \psi(k)$. As in Sections~\ref{sec:ex1} and \ref{sec:ex2}, \eqref{eq:single_psk} describes a thresholding operator over the $\sfM$-PSK constellation. Here, by growth of $\lambda$, the threshold $\tau_d$ increases,~and~consequently,~the~number~of~active transmit antennas reduces.
\begin{figure}[t]
\hspace*{-1.3cm}  
\resizebox{1.25\linewidth}{!}{
\pstool[width=.35\linewidth]{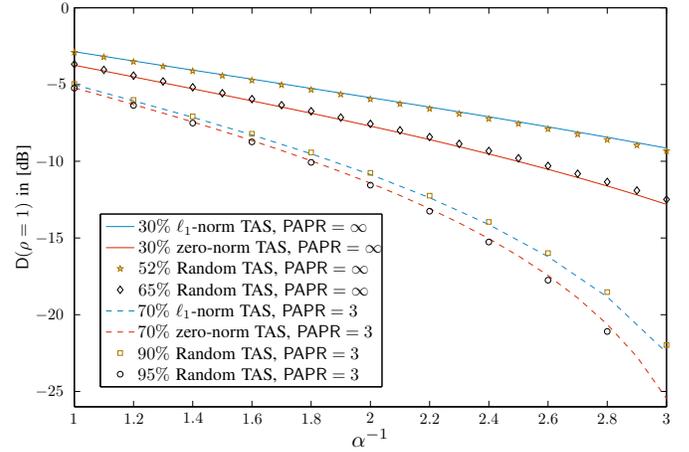}{
\psfrag{D}[c][c][0.2]{$\sfD(\rho=1)$ in [dB]}
\psfrag{alpha}[c][c][0.25]{$\alpha^{-1}$}
\psfrag{INFdb-normxx1-xx30XXXDDDXXXDDD}[l][l][0.2]{$30\%$ $\ell_1$-norm \ac{tas}, $\papr=\infty$}
\psfrag{INFdb-normxx0-xx30XXXDDDXXXDDD}[l][l][0.2]{$30\%$ zero-norm \ac{tas}, $\papr=\infty$}
\psfrag{INFdb-random1-052XXXDDDXXXDDD}[l][l][0.2]{$52\%$ Random \ac{tas}, $\papr=\infty$}
\psfrag{INFdb-random1-065XXXDDDXXXDDD}[l][l][0.2]{$65\%$ Random \ac{tas}, $\papr=\infty$}
\psfrag{03xdb-normxx1-xx70XXXDDDXXXDDD}[l][l][0.2]{$70\%$ $\ell_1$-norm \ac{tas}, $\papr=3$}
\psfrag{03xdb-normxx0-xx70XXXDDDXXXDDD}[l][l][0.2]{$70\%$ zero-norm \ac{tas}, $\papr=3$}
\psfrag{03xdb-random1-090XXXDDDXXXDDD}[l][l][0.2]{$90\%$ Random \ac{tas}, $\papr=3$}
\psfrag{03xdb-random1-095XXXDDDXXXDDD}[l][l][0.2]{$95\%$ Random \ac{tas}, $\papr=3$}
\psfrag{-5}[r][c][0.18]{$-5$}
\psfrag{-10}[r][c][0.18]{$-10$}
\psfrag{-15}[r][c][0.18]{$-15$}
\psfrag{-20}[r][c][0.18]{$-20$}
\psfrag{-25}[r][c][0.18]{$-25$}
\psfrag{-30}[r][c][0.18]{$-30$}
\psfrag{0}[r][c][0.18]{$0$}
%
\psfrag{1}[c][b][0.18]{$1$}
\psfrag{1.2}[c][b][0.18]{$1.2$}
\psfrag{1.4}[c][b][0.18]{$1.4$}
\psfrag{1.6}[c][b][0.18]{$1.6$}
\psfrag{1.8}[c][b][0.18]{$1.8$}
\psfrag{2}[c][b][0.18]{$2$}
\psfrag{2.2}[c][b][0.18]{$2.2$}
\psfrag{2.4}[c][b][0.18]{$2.4$}
\psfrag{2.6}[c][b][0.18]{$2.6$}
\psfrag{2.8}[c][b][0.18]{$2.8$}
\psfrag{3}[c][b][0.18]{$3$}
}}
\caption{\ac{rs}-predicted $\sfD(\rho)$ vs. $\alpha^{-1}$ for $\sfP=0.5$ considering no \ac{papr} limitation and $\papr=3$ dB. The zero-norm and $\ell_1$-norm precoders save $35\%$ and $22\%$ of active antennas in case of no \ac{papr} restriction, and about $25\%$ and $20\%$ when $\papr=3$ dB, respectively.\vspace*{-4mm}}
\label{fig:1}
\end{figure}
\subsection{Numerical Results}
Throughout the numerical investigations, the asymptotic fraction of active antennas is denoted by $\sfa$ which is determined by $\sfa=\E\mone\set{\xx^\dec(s^\dec)\neq 0}$ with $\mone\set{\cdot}$ being the indicator function. The average transmit power is represented by $\sfP$, and the \ac{papr} is denoted by $\papr$ which reads $\papr={\rmP}/\sfP$. We consider $\mH$ to be a fading channel whose entries~are~\ac{iid} with zero mean and variance $1/n$; thus, $\rmp_\mD$ follows Marcenko-Pastur's law, and $\rmR_\mD(\omega) = {\alpha}{(1-\omega)^{-1}}$ \cite{marvcenko1967distribution}.

Considering Sections \ref{sec:ex1} and \ref{sec:ex2}, Fig.~\ref{fig:1} shows the \ac{rs} predicted asymptotic distortion at $\rho=1$ in terms of the inverse load factor for two cases of $\papr=3$ dB and no peak power constraint. In the \ac{papr}-limited case, the curves have been sketched for $\sfa=0.7$, and in the other case, $\sfa=0.3$ has been considered; moreover, the average transmit power is set to be $\sfP=0.5$. As a benchmark, we have also plotted the points for random \ac{tas} which meet the corresponding curves. In fact, in the random \ac{tas}, the precoder selects a subset of transmit antennas randomly and precodes $\bs$ using the penalty function $u(\bv)=\lambda\norm{\bv}^2$. As the figure depicts, for the case of no peak power restriction, the zero-norm and $\ell_1$-norm based precoders need respectively about $35\%$ and $22\%$ fewer active transmit antennas compared to the random \ac{tas}. The gains in the case of $\papr=3$ dB reduce to $25\%$ and $20\%$ respectively.

\begin{figure}[t]
\hspace*{-1.25cm}  
\resizebox{1.23\linewidth}{!}{
\pstool[width=.35\linewidth]{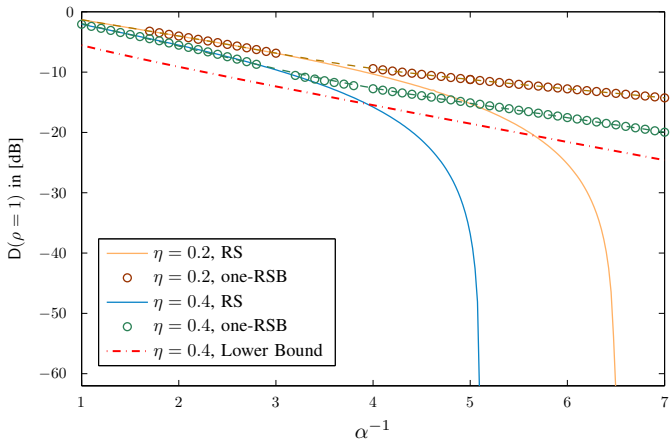}{
\psfrag{D}[c][b][0.2]{$\sfD(\rho=1)$ in [dB]}
\psfrag{alpha}[c][c][0.25]{$\alpha^{-1}$}
\psfrag{BPSK-a=0.2-N-RRS}[l][l][0.21]{$\sfa=0.2$, \ac{rs}}
\psfrag{BPSK-a=0.4-N-RRS}[l][l][0.21]{$\sfa=0.4$, \ac{rs}}
\psfrag{BPSK-a=0.2-N-RSB}[l][l][0.21]{$\sfa=0.2$, one-\ac{rsb}}
\psfrag{BPSK-a=0.4-N-RSB}[l][l][0.21]{$\sfa=0.4$, one-\ac{rsb}}
\psfrag{BPSK-a=0.2-L-RRS}[l][l][0.21]{$\sfa=0.4$, Lower Bound}


\psfrag{0}[r][c][0.18]{$0$}
\psfrag{-10}[r][c][0.18]{$-10$}
\psfrag{-20}[r][c][0.18]{$-20$}
\psfrag{-30}[r][c][0.18]{$-30$}
\psfrag{-40}[r][c][0.18]{$-40$}
\psfrag{-50}[r][c][0.18]{$-50$}
\psfrag{-60}[r][c][0.18]{$-60$}
\psfrag{-70}[r][c][0.18]{$-70$}
%
\psfrag{1}[c][b][0.18]{$1$}
\psfrag{2}[c][b][0.18]{$2$}
\psfrag{3}[c][b][0.18]{$3$}
\psfrag{4}[c][b][0.18]{$4$}
\psfrag{5}[c][b][0.18]{$5$}
\psfrag{6}[c][b][0.18]{$6$}
\psfrag{7}[c][b][0.18]{$7$}

}}
\caption{\ac{rs}- and one-step \ac{rsb}-predicted $\sfD(\rho)$ for BPSK signals with $\rmP=1$ under \ac{tas}. As $\alpha^{-1}$ grows, \ac{rs} violates the lower bound. The \ac{rsb} ansatz, however, is consistent with the lower bound.\vspace{-4mm}}
\label{fig:2}
\end{figure}

In order to investigate the impact of \ac{rsb}, we have also considered an example of antenna selection with BPSK transmission, i.e., $\sfM=2$ in Section \ref{sec:ex3}. Fig. \ref{fig:2} illustrates the \ac{rs} as well as one-step \ac{rsb} prediction of the asymptotic distortion at $\rho=1$ for two cases of $\sfa=0.2$ and $\sfa=0.4$ when $\rmP=1$. For sake of comparison, a theoretically rigorous lower bound for the case of $\sfa=0.4$ has been also sketched. The lower bound is derived as in \cite[Appendix C]{sedaghat2017lse}. As the figure shows, the \ac{rs} ansatz starts to fail predicting the asymptotic distortion as $\alpha^{-1}$ grows, and it even violates the lower bound in large inverse load factors. For this regime of $\alpha^{-1}$, however, the one-step \ac{rsb} ansatz gives a theoretically valid prediction.
\section*{Appendix: Large-System Analysis}
\label{sec:large}
In the sequel, we briefly sketch the derivations. Consider the Hamiltonian $\mae(\bv|\bs,\mH)=\norm{\mH \bv-\sqrt{\rho}\hspace*{.5mm}\bs}^2 +u(\bv)$, and define the partition function $\maz(\beta,h)$ to be
\begin{align}
\maz(\beta,h) =\sum_{\bv} e^{-\beta\mae(\bv|\bs,\mH)+hn\sfM^{\setW}_f(\bv;n)}.
\end{align}
By a standard large deviation argument, it is shown that 
\begin{align}
\sfM^{\setW}_f(\bx) =\lim_{n\uparrow\infty}  \lim_{\beta\uparrow\infty}  \frac{\partial}{\partial h} \maf(\beta,h)|_{h=0}, \label{eq:10}
\end{align}
in which $\maf(\beta,h) \coloneqq {n}^{-1}  \E \log \maz(\beta,h)$.
Moreover, the asymptotic distortion reads $\alpha \sfD (\rho) + \sfM_u^\setT (\bx) = \tilde{\mae}$ where we define $\setT(n) \coloneqq [1:n]$, and $\tilde{\mae}={\lim_{n\uparrow\infty} {n}^{-1} \E \mae(\bx|\bs,\mH)}$.~$\sfM_u^\setT (\bx)$ is determined in terms of $\maf(\cdot)$ by setting $f(x)\hspace*{-.7mm}=\hspace*{-.7mm}u(x)$ in \eqref{eq:10},~and
\begin{align}
\tilde{\mae} = - \lim_{n\uparrow\infty}  \lim_{\beta\uparrow\infty}  \frac{\partial}{\partial \beta} \maf(\beta,h)|_{h=0}. \label{eq:14}
\end{align}
Thus, the evaluation of $\sfD (\rho)$ and $\sfM_f^\setT (\bx)$ reduce to determining $\maf(\cdot)$; the task which we do via the replica method. 
Using the Riesz equality which states $ \E \log \xx = \lim\limits_{m\downarrow 0} {m}^{-1} \log \E \xx^m$,
\begin{align}
\maf(\beta,h) = \frac{1}{n} \lim_{m\downarrow 0}  \frac{1}{m} \log \E \left[ \maz(\beta,h) \right]^m. \label{eq:18}
\end{align}
\textit{Replica Method:} Evaluating $\maf(\beta,h)$ from \eqref{eq:18} is not trivial, as $m \in \setR^+$. The replica method determines the \ac{rhs} of \eqref{eq:18} by conjecturing the replica continuity. The replica continuity indicates that the ``analytic continuation'' of the non-negative integer moment function, i.e., $\E \left[ \maz(\beta,h) \right]^m$ for $m\in\setZ^+$, onto $\setR^+$ equals to the non-negative real moment function, i.e., $\E \left[ \maz(\beta,h) \right]^m$ for $m\in\setR^+$. The rigorous justification of the replica continuity has not been yet precisely addressed; however, the analytic results from the theory of spin glasses confirm the validity of the conjecture for several cases. 

Considering the replica continuity assumption, Proposition~\ref{thm:1} is concluded by taking some lines of calculations form \eqref{eq:18} which have been left for the extended version of~the~manus- cript due to the page limitation.\vspace*{2mm}

\bibliography{ref}
\bibliographystyle{IEEEtran}
\begin{acronym}
\acro{mimo}[MIMO]{Multiple-Input Multiple-Output}
\acro{csi}[CSI]{Channel State Information}
\acro{awgn}[AWGN]{Additive White Gaussian Noise}
\acro{iid}[i.i.d.]{independent and identically distributed}
\acro{ut}[UT]{User Terminal}
\acro{bs}[BS]{Base Station}
\acro{tas}[TAS]{Transmit Antenna Selection}
\acro{lse}[LSE]{Least Square Error}
\acro{rhs}[r.h.s.]{right hand side}
\acro{lhs}[l.h.s.]{left hand side}
\acro{wrt}[w.r.t.]{with respect to}
\acro{rs}[RS]{Replica Symmetry}
\acro{rsb}[RSB]{Replica Symmetry Breaking}
\acro{papr}[PAPR]{Peak-to-Average Power Ratio}
\acro{rzf}[RZF]{Regularized Zero Forcing}
\acro{snr}[SNR]{Signal-to-Noise Ratio}
\acro{rf}[RF]{Radio Frequency}
\acro{mf}[MF]{Match Filtering}
\end{acronym}
\end{document}

%% file: commands.tex
\newcommand{\setX}{\mathbbmss{X}}

\newcommand{\setR}{\mathbbmss{R}}

\newcommand{\setW}{\mathbbmss{W}}

\newcommand{\setZ}{\mathbbmss{Z}}

\newcommand{\setC}{\mathbbmss{C}}

\newcommand{\setT}{\mathbbmss{T}}

\newcommand{\rmP}{\mathrm{P}}

\newcommand{\rmp}{\mathrm{p}}

\newcommand{\rmR}{\mathrm{R}}

\newcommand{\rmG}{\mathrm{G}}

\newcommand{\rmj}{\mathrm{j}}

\newcommand{\dec}{\mathsf{dec}}
\newcommand{\sfP}{\mathsf{P}}

\newcommand{\her}{\mathsf{H}}
\newcommand{\sfc}{\mathsf{c}}

\newcommand{\sfM}{\mathsf{M}}

\newcommand{\sfD}{\mathsf{D}}

\newcommand{\sfR}{\mathsf{R}}
\newcommand{\sfd}{\mathsf{d}}

\newcommand{\sfK}{\mathsf{K}}
\newcommand{\sfp}{\mathsf{p}}
\newcommand{\sfa}{\eta}

\newcommand{\maf}{\mathcal{F}}

\newcommand{\mae}{\mathcal{E}}
\newcommand{\maz}{\mathcal{Z}}

\newcommand{\mad}{\mathcal{D}}

\newcommand{\mai}{\mathcal{I}}

\newcommand{\bvv}{\mathbf{v}}

\newcommand{\bx}{{\boldsymbol{x}}}

\newcommand{\xx}{\mathrm{x}}

\newcommand{\set}[1]{\left\lbrace#1\right\rbrace}

\newcommand{\bz}{{\boldsymbol{z}}}

\newcommand{\bs}{{\boldsymbol{s}}}

\newcommand{\bv}{{\boldsymbol{v}}}

\newcommand{\dif}{\mathrm{d}}

\newcommand{\by}{{\boldsymbol{y}}}

\newcommand{\mI}{\mathbf{I}}
\newcommand{\mone}{\mathbf{1}}

\newcommand{\mQ}{\mathbf{Q}}

\newcommand{\mU}{\mathbf{U}}
\newcommand{\mD}{\mathbf{D}}
\newcommand{\mM}{\mathbf{M}}

\newcommand{\mT}{\mathbf{T}}
\newcommand{\mH}{\mathbf{H}}

\newcommand{\E}{\mathsf{E}\hspace{.5mm}}

\newcommand{\rs}{{\mathsf{rs}}}
\newcommand{\rsb}{{\mathsf{rsb}}}
\newcommand{\papr}{{\mathsf{PAPR}}}

\newcommand{\norm}[1]{\lVert #1 \rVert}
\newcommand{\re}[1]{\mathsf{Re}\left\lbrace #1 \right\rbrace}

\newcommand{\abs}[1]{\lvert #1 \rvert}
\newcommand{\tr}[1]{\mathrm{Tr} \left\lbrace #1 \right\rbrace}

\newtheoremstyle{mystyle}
  {}
  {}
  {}
  {}
  {\bfseries}
  {:}
  { }
  {}

\theoremstyle{mystyle}

\newtheorem{definition}{Definition}
\newtheorem{proposition}{Proposition}
\newtheorem*{remark}{Remark}
\newtheorem*{prf}{Proof}
\newtheorem{corollary}{Corollary}
\newtheorem*{decoupling}{General Marginal Decoupling Property}
\newtheorem*{decouplersb}{RSB Marginal Decoupling Property}
\newcounter{bar}

